\documentclass[12pt,a4paper]{article}
\usepackage{latexsym,amssymb,amsmath,amsthm}
\usepackage[cp1251]{inputenc}

\newtheorem{theorem}{Theorem}

\newtheorem{corol}{Corollary}

\newcommand{\OR}{\mathsf{OR}}
\newcommand{\XOR}{\mathsf{XOR}}
\newcommand{\SUM}{\mathsf{SUM}}
\newcommand{\LL}{\mathsf{L}}

\newenvironment{pnote}{\smallskip \small \noindent $\blacktriangleright$}{\hfill$\blacktriangleleft$\smallskip}

\begin{document}
\title{Notes on the complexity of coverings for Kronecker powers of symmetric matrices}
\date{}
\author{Igor S. Sergeev\footnote{e-mail: isserg@gmail.com}}

\maketitle

\begin{abstract}
In the present note, we study a new method of constructing
efficient coverings for Kronecker powers of matrices, recently
proposed by J.~Alman, Y.~Guan, A.~Padaki~\cite{agp22e}. We provide
an alternative proof for the case of symmetric matrices in a stronger form.
As a consequence, the previously known upper bound on the depth-2 additive
complexity of the boolean $N\times N$ Kneser-Sierpinski matrices is improved to $O(N^{1.251})$.
This work can be viewed as a supplement~to~\cite{js13e}.
\end{abstract}

\section{Introduction}

Let us recall necessary concepts. See~\cite{js13e} for a more detailed introduction to the subject.

A {\it rectangle} of size $a \times b$ is an all-1s matrix with
$a$ rows and $b$ columns. Further, depending on the context, sometimes under rectangle we will understand
a rank-1 matrix, i.e. consisting of an all-1s submatrix and all 0s in other entries.

%The {\it weight} of a matrix $A$ is defined as the number of its nonzero elements, and is denoted as $|A|$.
%For an $a \times b$ rectangle, we have $|R| = ab$.
We define the {\it complexity}\footnote{In~\cite{js13e}, we used a
term {\it weight} instead. Here we substitute it with complexity
to avoid confusing with the spectral weight of a rectangle.} of an
$a \times b$ rectangle $R$ as the sum of lengths of its two sides,
$w(R) = a+b$. We introduce the characteristic of the narrowness of
a rectangle as the ratio of the lengths of its larger and smaller
sides, $\rho(R) = \frac{\max(a,\,b)}{\min(a,\,b)}$. The  {\it
spectral weight} of a rectangle is defined as $\sigma(R) =
\sqrt{ab}$.

A set $F = \{ R_1, \ldots, R_k \}$ of rectangles is a {\it covering} of a boolean matrix~$A$, if
\begin{equation}\label{cov}
 A = R_1 + \ldots + R_k.
\end{equation}
(Here under $R_i$ we mean rank-1 matrices.) If the operation ``+''
in~(\ref{cov}) is an integer addition, then $F$ is called
$\SUM$-covering. If ``+'' is a disjunction, then $F$ is called
$\OR$-covering. In the case when ``+'' is a $\bmod\,2$ addition,
then we have an $\XOR$-covering.

The complexity of a covering $F$ is defined as $w(F) = w(R_1) +
\ldots + w(R_k)$, and the spectral weight as $\sigma(F) =
\sigma(R_1) + \ldots + \sigma(R_k)$. The {\it $\LL$-complexity} of
a matrix $A$ is defined for $\LL \in \{\SUM, \OR, \XOR \}$ as the
minimal complexity of its $\LL$-covering, denoted by $\LL_2(A)$
(it means the complexity of computation of~$A$ by depth-2 linear
circuits of the corresponding type).

\begin{pnote}
These notions may be extended to the case of matrices over an arbitrary semiring~$S$.
A rectangle over $S$ is a matrix $(c_1, \ldots, c_a)^T \cdot (d_1, \ldots, d_b)$, where $c_i, d_j \in
S \setminus \{0\}$. A covering of a matrix is conditioned by
\[  A = e_1 R_1 + \ldots + e_k R_k, \qquad e_i \in S.  \]
The results presented below may be applied also to the analogously
defined measure of complexity of computation of matrices by
algebraic linear circuits of depth~2.
\end{pnote}

Let $\sigma(A)$ denote the minimal spectral weight of a matrix
$A$. Since $w(R) \ge 2\sigma(R)$ for any rectangle $R$, spectral
weight serves as a simple lower bound for complexity\footnote{Here
and below, symbols $\asymp$, $\prec$, $\preceq$ denote the
equality, strict and non-strict inequalities on the order of
growth.}: $\LL_2(A) \succeq \sigma(A)$.

A convenient property of the spectral weight is its
multiplicativity with respect to the Kronecker product. Recall
that the  {\it Kronecker product} of boolean matrices $A$, $B$ is
a matrix $A \otimes B$ obtained by replacing 1-entries of~$A$ by
copies of $B$, and 0-entries by all-0s matrices of the same size.

Note that if $F$ and $G$ are coverings of matrices $A$ and $B$, then $F
\otimes G$ is a covering\footnote{The Kronecker product of sets of matrices is
 $F \otimes G = \{ R \otimes R' \mid R \in F, \, R' \in G \}$.} of $A \otimes B$, and $\sigma(F
\otimes G) = \sigma(F)\sigma(G)$. In particular, if we construct a
covering of a matrix $A^{\otimes n}$ (a Kronecker power of $A$) by
the product-of-coverings method above using appropriate coverings
of $A$, then the complexity of a resulting covering $H$ satisfies
$w(H) \succeq \sigma(H) \succeq \sigma^n(A)$.

In~\cite{agp22e}, the authors actually pose a question: can we
obtain upper bounds like $\LL_2(A^{\otimes n}) \preceq
\sigma^{n+o(n)}(A)$ or at least $\LL_2(A^{\otimes n}) \preceq
\sigma^{n+o(n)}(F)$ for some appropriate coverings $F$ of a matrix
$A$. In general, it is not possible, just consider an example $A = \begin{bmatrix} 1 & 1 \\ 0 & 0 \end{bmatrix}$.

A general obstacle for the desired bounds is the growing narrowness
of the covering rectangles. Note that the complexity and the
spectral weight of a rectangle $R$ are related as\footnote{To be
precise, $w(R) = \left(\sqrt{\rho(R)}
+\sqrt{1/\rho(R)}\right)\sigma(R)$.} $w(R) \asymp
\sqrt{\rho(R)}\sigma(R)$. Nevertheless, under some conditions, the
authors of~\cite{agp22e} were able to overcome the indicated
obstacle and to derive the desired bounds. The first situation is
when the covering $F$ satisfies some asymmetry criteria, the
second one is when the matrix $A$ is symmetric, and the covering
$F$ is one-sided (it means that all rectangles are stretched in
the same direction).

Perhaps, the most challenging object (among boolean matrices) to
apply the theory are the Kneser--Sierpinski (or disjointness)
matrices. Recall that the boolean $N \times N$ {\it
Kneser--Sierpinski matrix} $D_N$ is defined for $N=2^n$ as
follows. Rows and columns of $D_N$ are labeled by distinct subsets
$u \in [n]$, and $D[u,v] = 1$ iff $u \cap v = \emptyset$. A matrix
$D_N$ also may be viewed as a Kronecker product
\begin{equation}\label{D}
D_N = D_2^{\otimes n} = \underbrace{D_2 \otimes \ldots \otimes
D_2}_n,  \qquad
   D_2 = \begin{bmatrix} 1 & 1 \\ 1 & 0 \end{bmatrix}.
\end{equation}

The problem of complexity of $\OR$-coverings for $D_N$ was almost
closed in~\cite{cils17e}. There was established that
\[ N^{1.16} \prec \OR_2(D_N) \prec N^{1.17} \]
(the lower bound is from~\cite{js13e}). Moreover, the authors
of~\cite{cils17e} constructed a covering of almost minimal
complexity, up to a factor of order $(\log N)^{O(1)}$.

The question about additive ($\SUM$) complexity of matrices is
less clear. In~\cite{js13e}, we propose a simple way to show that
$$\SUM_2(D_N) \preceq \sigma^n(D_2) = \left(\sqrt2+1\right)^n \prec N^{1.272}$$
relying on a trivial (and optimal) decomposition of the matrix
$D_2$ into rectangles of size $1 \times 1$ and $1\times 2$ (or
$2\times 1$). This approach was nontrivially generalized
in~\cite{agp22e}. Due to limitations inherent in the analysis of
the proposed method, the efficient implementation of the
Kneser--Sierpinski matrices is justified only with the basic
coverings of matrices $D_4$ and $D_8$. In the latter case, the
obtained bound~\cite{agp22e} is
$$\SUM_2(D_N) \preceq \sigma^{n/3+o(n)}(D_8) = \left(\sqrt8+\sqrt7+3\sqrt3+3\right)^{n/3+o(n)} \prec N^{1.258}.$$

In the present note, we describe a version of this method for the
(most interesting) case of symmetric matrices. The limits of
applicability of the method are (comparatively) extended, and a
more elementary proof is given. As a consequence, an upper bound
for the complexity of the Kneser--Sierpinski matrices is reduced
to $\SUM_2(D_N) \prec N^{1.251}$.

Note that the question about existence of substantially more
efficient $\XOR$-coverings for matrices $D_N$ is still open.

\section{The synthesis method}

In this section, we provide a general method for constructing a
covering of a symmetric matrix $A^{\otimes n}$. The method is
essentially equivalent to the method~\cite{agp22e}. By the way, we
will follow an illustrating example where a non-trivial covering
of $D_N$ is obtained from coverings of the matrix $D_4$. This
example doesn't require the method in its full generality.

Assume there exist two coverings of a symmetric matrix $A$, or to
be precise, two pairs of coverings, if we count transposed ones.
The first covering $F$ ($F^T$) is supposed to be efficient in the
terms of spectral weight. The second covering $G$ is one-sided:
the longest sides of all its non-square rectangles are parallel.
Coverings $G/G^T$ serve to compensate an imbalance caused by the
use of $F$-type coverings.

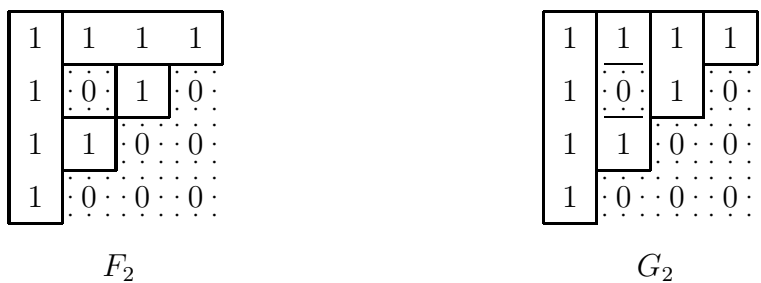
\begin{figure}[htb]
\begin{center}
\begin{picture}(280,100)(0,0)

\thicklines

\multiput(0,20)(20,0){2}{\line(0,1){80}}
\multiput(0,20)(0,80){2}{\line(1,0){20}}
\multiput(20,80)(0,20){2}{\line(1,0){60}}
\multiput(40,40)(20,20){3}{\line(0,1){20}}
\put(20,40){\line(1,0){20}} \put(20,60){\line(1,0){40}}
\put(40,60){\line(0,1){20}}

\multiput(7,26)(0,20){4}{$1$} \multiput(27,86)(20,0){3}{$1$}
\multiput(27,46)(20,20){2}{$1$}

\multiput(27,26)(20,0){3}{$0$} \multiput(47,46)(20,0){2}{$0$}
\multiput(27,66)(40,0){2}{$0$}

\put(35,0){$F_2$}

\multiput(200,20)(20,0){2}{\line(0,1){80}}
\multiput(200,20)(20,20){4}{\line(1,0){20}}
\put(200,100){\line(1,0){80}} \put(240,40){\line(0,1){60}}
\put(260,60){\line(0,1){40}} \put(280,80){\line(0,1){20}}

\multiput(207,26)(0,20){4}{$1$} \multiput(227,86)(20,0){3}{$1$}
\multiput(227,46)(20,20){2}{$1$}

\multiput(227,26)(20,0){3}{$0$} \multiput(247,46)(20,0){2}{$0$}
\multiput(227,66)(40,0){2}{$0$}

\put(235,0){$G_2$}

\thinlines

\multiput(223,60)(0,20){2}{\line(1,0){14}}
%\put(220,60){\line(1,1){20}} \put(220,80){\line(1,-1){20}}

\multiput(23,23)(0,7){3}{\circle*{1}}
\multiput(37,23)(0,7){3}{\circle*{1}}
\multiput(30,23)(0,14){2}{\circle*{1}}
\multiput(23,63)(0,7){3}{\circle*{1}}
\multiput(37,63)(0,7){3}{\circle*{1}}
\multiput(30,63)(0,14){2}{\circle*{1}}
\multiput(43,23)(0,7){3}{\circle*{1}}
\multiput(57,23)(0,7){3}{\circle*{1}}
\multiput(50,23)(0,14){2}{\circle*{1}}
\multiput(43,43)(0,7){3}{\circle*{1}}
\multiput(57,43)(0,7){3}{\circle*{1}}
\multiput(50,43)(0,14){2}{\circle*{1}}
\multiput(63,43)(0,7){3}{\circle*{1}}
\multiput(77,43)(0,7){3}{\circle*{1}}
\multiput(70,43)(0,14){2}{\circle*{1}}
\multiput(63,23)(0,7){3}{\circle*{1}}
\multiput(77,23)(0,7){3}{\circle*{1}}
\multiput(70,23)(0,14){2}{\circle*{1}}
\multiput(63,63)(0,7){3}{\circle*{1}}
\multiput(77,63)(0,7){3}{\circle*{1}}
\multiput(70,63)(0,14){2}{\circle*{1}}

\multiput(223,23)(0,7){3}{\circle*{1}}
\multiput(237,23)(0,7){3}{\circle*{1}}
\multiput(230,23)(0,14){2}{\circle*{1}}
\multiput(223,63)(0,7){3}{\circle*{1}}
\multiput(237,63)(0,7){3}{\circle*{1}}
\multiput(230,63)(0,14){2}{\circle*{1}}
\multiput(243,23)(0,7){3}{\circle*{1}}
\multiput(257,23)(0,7){3}{\circle*{1}}
\multiput(250,23)(0,14){2}{\circle*{1}}
\multiput(243,43)(0,7){3}{\circle*{1}}
\multiput(257,43)(0,7){3}{\circle*{1}}
\multiput(250,43)(0,14){2}{\circle*{1}}
\multiput(263,43)(0,7){3}{\circle*{1}}
\multiput(277,43)(0,7){3}{\circle*{1}}
\multiput(270,43)(0,14){2}{\circle*{1}}
\multiput(263,23)(0,7){3}{\circle*{1}}
\multiput(277,23)(0,7){3}{\circle*{1}}
\multiput(270,23)(0,14){2}{\circle*{1}}
\multiput(263,63)(0,7){3}{\circle*{1}}
\multiput(277,63)(0,7){3}{\circle*{1}}
\multiput(270,63)(0,14){2}{\circle*{1}}

\end{picture}
\caption{Weight-minimal and compensating coverings of $D_4$}\label{pic_cov}
\end{center}
\end{figure}

\begin{pnote}
Fig.~\ref{pic_cov} shows appropriate coverings of $D_4$. The
covering $F_2$ (on the left) has optimal spectral weight
$\sigma(F_2) = 4+\sqrt3$. The covering $G_2$ (on the right) has
good correcting qualities. Its spectral weight is slightly higher,
$\sigma(G_2)=3+2\sqrt2$.
\end{pnote}

Let $H$ be a covering of a matrix $B$. Then we can build a
covering of a matrix $A \otimes B$ in the form $\{ F_i \otimes R_i
\mid R_i \in H \}$, where $F_i$ are some coverings of~$A$. This
way we sequentially obtain coverings for matrices $A,
A^{\otimes2}, A^{\otimes3}, \ldots$ In the main process, we choose
$F_i \in \{ F,\, F^T \}$. Precisely, we transform an $a \times b$
rectangle $R$ into $F\otimes R$, if $a \le b$, and into $F^T
\otimes R$, otherwise.

Consider a covering $F = \{ R_1, \ldots, R_s \}$ consisting of $a_i \times b_i$ rectangles~$R_i$. The {\it
characteristic function} of $F$ is defined as
\begin{equation}\label{char}
    \chi_F(x) = \sigma(R_1)\left( \frac{a_1}{b_1} \right)^x + \ldots + \sigma(R_s)\left( \frac{a_s}{b_s} \right)^x - \sigma(F).
\end{equation}
As follows from the definition, $\chi_F(0)=0$. We call a covering $F$ {\it
compact}, if $\chi_F(x)$ takes negative values on some negative arguments\footnote{It is a weak analogue
of imbalanced covering from~\cite{agp22e}.}. For a compact covering $F$,
let $\lambda_F$ denote the minimal real root of $\chi_F$ in whose right semineighbourhood
the function is negative\footnote{The compactness of a covering implies that $b_i <
a_i$ for some $i$. Then, $\lambda_F$ is correctly defined since $\chi_F(x) \to +\infty$ as $x \to
-\infty$, and any exponential sum of the form~(\ref{char}) has a finite number of real zeros, see e.g.~\cite{lan31e}.}.

Note that the condition $\chi'_F(0) = \sum_i \sigma(R_i) \ln(a_i/b_i)
> 0$ is sufficient for~$F$ to be compact.

So, we require the compactness of a covering $F$ for our algorithm of computation
of a Kronecker power of a matrix $A$.

\begin{pnote}
The covering $F_2$ of $D_4$ is compact. Its characteristic function
is $\chi_{F_2}(x) = 2\cdot4^x + \sqrt3\cdot3^{-x} - 2- \sqrt{3}$, and the minimal root is
$\lambda_{F_2} \approx -0.305$.
\end{pnote}

For compensation, we will use a compact {\it one-sided} covering
$G$ consisting of $a'_i \times b'_i$ rectangles $R'_i$ satisfying
$a'_i \ge b'_i$. The quality of such covering is characterized by
the coefficient
\[  \mu_G = \frac1{\sigma(G)} \sum_{R \in G} \frac{\sigma(R)}{\sqrt{\rho(R)}} = \frac1{\sigma(G)} \sum_{i} b'_i.  \]

For $x>1$, define the function
\begin{equation}\label{pi_G}
  \pi_G(x) = \frac1{\sigma(G)} \sum_{R \in G} \sigma(R)\cdot x^{-\lfloor \log_x \rho(R) \rfloor/2}.
\end{equation}
It easily follows from the definition that $\pi_G(x) \ge \mu_G$, and $\pi_G(x) \to \mu_G$ as $x \to 1$.

\begin{pnote}
For the covering $G_2$, we have $\mu_{G_2} = \frac4{3+2\sqrt2}$.
\end{pnote}

\begin{theorem}\label{main}
Let $F$ be a compact $\LL$-covering, and $G$ be a compact one-sided $\LL$-covering of
a symmetric $r\times r$ matrix $A$, and $\sigma(G) \ge \sigma(F)$. If the condition
\begin{equation}\label{cond}
   \frac{\sigma(G)}{\sigma(F)} < \mu_G^{2\lambda_F}
\end{equation}
is satisfied, then for $N=r^n$,
\[ \LL_2(A^{\otimes n}) \preceq N^{\log_r \sigma(F)}. \]
\end{theorem}

\begin{pnote}
The matrix $D_4$ and its coverings $F_2, G_2$ satisfy the conditions of the theorem, since
$\lambda_{F_2} < -0.3$ and $\mu_{G_2}= \frac4{3+2\sqrt2}$.
Hence, $\SUM_2(D_N) \preceq N^{\log_4(4+\sqrt3)} \prec N^{1.26}.$
\end{pnote}

\proof The proof strategy is the following. First, we analyze the
evolution of rectangle sizes after multiple application of
type-$F$ coverings. Then, do the same for type-$G$ coverings.
Finally, we propose an appropriate combination of these two types
of coverings.

{\bf I.} By compactness of the covering $F$, for some (small
enough) $\delta, \epsilon >0$, and $\lambda = \lambda_F+\delta$
such that $\chi_F(\lambda) < - \epsilon$, the
inequality~(\ref{cond}) holds true after replacing $\lambda_F$ by
$\lambda$. Let us check that for all small enough $\tau>1$,
\begin{equation}\label{chi_app}
    \sigma(R_1)\tau^{\lambda \lfloor \log_{\tau}(a_1/b_1) \rfloor} + \ldots +
    \sigma(R_s)\tau^{\lambda \lfloor \log_{\tau}(a_s/b_s) \rfloor} \le \sigma(F).
\end{equation}
Indeed, the left side of (\ref{chi_app}) doesn't exceed
\[ \tau^{-\lambda}(\chi_F(\lambda)+\sigma(F)) < \tau^{-\lambda}(\sigma(F)-\epsilon), \]
thus to satisfy (\ref{chi_app}), it is sufficient to require
$\tau^{-\lambda} \le 1 + \frac{\epsilon}{\sigma(F)}$. Therefore,
any choice from the interval $1 < \tau \le \left( 1 +
\frac{\epsilon}{\sigma(F)} \right)^{-1/\lambda}$ is suitable.

We assign to the parameter $\tau$ the meaning of a discretization
step of changing the ratios between the rectangle's sides in the
classification of rectangles. The final choice of $\tau$ will be
decided later.

Let us introduce a classification on the set $\cal R$ of all
rectangles depending on the ratio between the longer and the
shorter sides. Set ${\cal R} = \bigcup_{k \ge 0} I_k$, where $I_0$
contains rectangles $R$ satisfying $\rho(R)\le r$, and for $k \ge
1$, the set $I_k$ contains rectangles with ratios
$r\cdot\tau^{k-1} < \rho(R)\le r\cdot\tau^{k}$. Recall that $r$ is
the size of the matrix $A$.

\begin{pnote}
In the example with the covering $F_2$, we set $\tau = 4$. Then, $I_0 = \{ R
\mid \rho(R) \le 4 \}$, and $I_k = \{ R \mid 4^k < \rho(R) \le
4^{k+1} \}$ for $k>0$. The function $\chi_{F_2}(x)$ is negative in the interval $(\lambda_{F_2},\, 0)$,
thus we are quite free in choosing $\lambda$. The specific value will be determined later.
\end{pnote}

When performing compositions with coverings of $A$, we will track
the distribution of the spectral weight of rectangles among the
sets $I_k$. In doing so, we will be guided by the principle of
{\it error to the right}. It means that: (a) we allow a rectangle
to be placed into a set $I_k$ with a higher index $k$, but not
otherwise, and (b) we estimate the distribution of the weight of a
covering of $A \otimes R$ only on the basis of information about
assigning $R$ to a certain set $I_k$. As a consequence, the
estimated distribution of a covering obtained as a result of a
series of iterations (with possible errors to the right) may
differ from the true distribution only in that some rectangles
appear in sets $I_k$ with higher indices.

The redistribution of the spectral weight of a set of rectangles
under the composition with type-$F$ coverings (with possible
errors to the right) may be estimated from the coefficients of the
Laurent polynomial
\begin{equation}\label{laur}
   P_F(x) = \sum_{i \in \mathbb Z} \beta_i x^i = \frac{\sigma(R_1)}{\sigma(F)} x^{\lfloor \log_{\tau}(a_1/b_1) \rfloor} + \ldots +
    \frac{\sigma(R_s)}{\sigma(F)} x^{\lfloor \log_{\tau}(a_s/b_s) \rfloor}
\end{equation}
obtained from~(\ref{chi_app}). It means that for $R \in I_m$, the
weight of rectangles from $(F/F^T)\otimes R$ is distributed so
that the portion $\beta_k$ of the total weight belongs to
$I_{m-k}$ in the case $m-k>0$, and to $I_0$, otherwise.

\begin{figure}[htb]
\begin{center}
\begin{picture}(320,60)(0,15)

\multiput(10,30)(50,0){3}{\circle*{3}} \put(250,30){\circle*{3}}
% right arrows
\multiput(44,34)(50,0){3}{\vector(4,-1){12}}
\multiput(14,31)(50,0){3}{\line(4,1){12}}
\multiput(234,34)(50,0){2}{\vector(4,-1){12}}
\multiput(204,31)(50,0){2}{\line(4,1){12}}
\qbezier[40](26,34)(35,36)(44,34)
\qbezier[40](76,34)(85,36)(94,34)
\qbezier[40](126,34)(135,36)(144,34)
\qbezier[40](216,34)(225,36)(234,34)
\qbezier[40](266,34)(275,36)(284,34)

% left arrows
\multiput(26,26)(50,0){3}{\vector(-4,1){12}}
\multiput(56,29)(50,0){3}{\line(-4,-1){12}}
\multiput(216,26)(50,0){2}{\vector(-4,1){12}}
\multiput(246,29)(50,0){2}{\line(-4,-1){12}}
\qbezier[40](26,26)(35,24)(44,26)
\qbezier[40](76,26)(85,24)(94,26)
\qbezier[40](126,26)(135,24)(144,26)
\qbezier[40](216,26)(225,24)(234,26)
\qbezier[40](266,26)(275,24)(284,26)

% cycles
\multiput(15,45)(50,0){3}{\vector(-1,-3){4}}
\put(255,45){\vector(-1,-3){4}}
\multiput(9,33)(50,0){3}{\line(-1,3){4}}
\put(249,33){\line(-1,3){4}}%
\qbezier[40](5,45)(1,55)(10,55)%
\qbezier[40](15,45)(19,55)(10,55)
\qbezier[40](55,45)(51,55)(60,55)
\qbezier[40](65,45)(69,55)(60,55)
\qbezier[40](105,45)(101,55)(110,55)
\qbezier[40](115,45)(119,55)(110,55)
\qbezier[40](245,45)(241,55)(250,55)
\qbezier[40](255,45)(259,55)(250,55)

% signatures
\put(6,15){$I_0$} \put(56,15){$I_1$} \put(106,15){$I_2$}
\put(246,15){$I_k$}

\multiput(28,17)(50,0){3}{$\scriptstyle{2/\omega}$}
\multiput(218,17)(50,0){2}{$\scriptstyle{2/\omega}$}
\multiput(23,39)(50,0){3}{$\scriptstyle{\sqrt3/\omega}$}
\multiput(213,39)(50,0){2}{$\scriptstyle{\sqrt3/\omega}$}
\multiput(53,59)(50,0){2}{$\scriptstyle{2/\omega}$}
\put(243,59){$\scriptstyle{2/\omega}$}
\put(3,59){$\scriptstyle{4/\omega}$}

\put(174,30){\ldots}  \put(314,30){\ldots}

\end{picture}
\caption{Diagram of spectral weight redistribution under the
action of the composition with the covering $F_2/F_2^T$ (here
$\omega=\sigma(F_2)=4+\sqrt3$)}\label{pic_spF}
\end{center}
\end{figure}
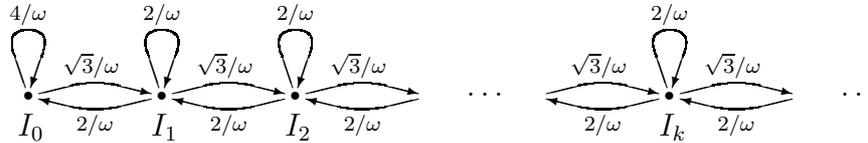

\begin{pnote}
For the covering $F_2$, and $\tau=4$, we obtain $P_{F_2}(x) =
\frac{2x + 2 + \sqrt3x^{-1} }{4+\sqrt3}$. The corresponding
redistribution diagram is shown on Fig.~\ref{pic_spF}.
\end{pnote}

Let $p_k(t)$ stand for the fraction of the spectral weight of the
constructed covering of $A^{\otimes t}$ associated with the set
$I_k$. In the beginning, one has $p_0(0)= 1$, and $p_k(0)=0$ for
all $k>0$. Set $\nu=\tau^{\lambda}$. By~(\ref{chi_app}),
$P_F(\nu)\le1$. Denote $d = \deg P_F = \max_{\beta_i >0} |i|$.

We are going to show that the distribution $\{ p_k^*(t) \}$ with
$p^*_k(t) = \nu^k$ for $k \le dt$, and $p^*_k(t) = 0$ for all $k >
dt$, is a majorant for $\{ p_k(t) \}$, meaning that the values
$p^*_k(t)$ upper bound the components of some distribution
$\{p'_k(t)\}$ obtained from $\{ p_k(t) \}$ by a partial shift of
the distribution to the right: from components with smaller
indices to components with greater indices\footnote{Though the
distribution $\{ p_k(t) \}$ is probabilistic, we don't require the
same from the majorant $\{ p_k^*(t) \}$ allowing the sum of its
components be greater than 1. Our goal is just deriving upper
bounds on $p_k(t)$.}.

Obviously, in the moment $t=0$, the majorization condition is
fulfilled. Let us prove the induction step: apply the composition
with $F/F^T$ to a set of rectangles with the distribution $\{
p_k^*(t)\}$. Rectangles from $I_k$, where $k > d(t+1)$, do not
appear here. For $0<k \le d(t+1)$, the weight of rectangles from
$I_k$ may be upper bounded as
\begin{equation}\label{w_bal}
     \sum_{i \in \mathbb Z} \beta_i \nu^{k+i} = \nu^k P_F(\nu) \le \nu^k.
\end{equation}
In the case $k=0$, this bound is, generally speaking, wrong, since
an essential portion of the total weight remains in $I_0$.

However, if $I_0$ receives exceptional weight, then other sets
$I_1, I_2, \ldots$ (in general) suffer from the weight deficit, as
follows from~(\ref{w_bal}). Therefore, we can redistribute the
exceptional weight from $I_0$ to other sets, in accordance to the
error-to-the-right principle. It is possible, since the
composition preserves the conditional total weight, i.e. the sum
of distribution components.

As a consequence, after $t$ steps of composition with coverings
$F/F^T$, any set $I_k$ contains a portion at most $\nu^k$ of the
total weight $\sigma^t(F)$ of the resulting covering of
$A^{\otimes t}$, up to some errors to the right.

\begin{pnote}
In our example with $D_4$, we may choose $\nu = \sqrt3/2$ (it's a
root of the polynomial $P_{F_2}(x)$). The choice allows us to
avoid a redistribution of the weight from~$I_0$. The distribution
$\{1, \nu, \nu^2, \ldots, \nu^k, \ldots\}$ is stationary for the
diagram on Fig.~\ref{pic_spF}. Implicitly, we also chose $\lambda
= \log_4\nu > \lambda_{F_2}$.
\end{pnote}

{\bf II.} The redistribution of the spectral weight of a set of
rectangles under the composition with type-$G$ coverings (again,
with possible errors to the right) may be described by the
polynomial
\[ P_G(x) = \alpha_l x^{l} + \ldots + \alpha_1 x + \alpha_0 = \frac1{\sigma(G)} \sum_{R \in G} \sigma(R)x^{\lfloor \log_{\tau} \rho(R) \rfloor}. \]
For $R \in I_m$, the spectral weight of rectangles from
$(G/G^T)\otimes R$ is distributed so that the portion $\alpha_k$
belongs to  $I_{m-k}$ in the case $k<m$, and to $I_0$, otherwise.

\begin{pnote}
Actually, for the covering $G_2$, and $\tau=4$, one can assign
(assuming some errors to the right) $P_{G_2}(x) = (x+2)/3$. The
weight redistribution under the action of coverings $G_2/G_2^T$ is
shown by the diagram on Fig.~\ref{pic_spG}.
\end{pnote}

\begin{figure}[htb]
\begin{center}
\begin{picture}(320,60)(0,15)

\multiput(10,30)(50,0){3}{\circle*{3}} \put(250,30){\circle*{3}}

% left arrows
\multiput(57,30)(50,0){3}{\vector(-1,0){44}}
\multiput(247,30)(50,0){2}{\vector(-1,0){44}}

% cycles
\multiput(15,45)(50,0){3}{\vector(-1,-3){4}}
\put(255,45){\vector(-1,-3){4}}
\multiput(9,33)(50,0){3}{\line(-1,3){4}}
\put(249,33){\line(-1,3){4}}%
\qbezier[40](5,45)(1,55)(10,55)%
\qbezier[40](15,45)(19,55)(10,55)
\qbezier[40](55,45)(51,55)(60,55)
\qbezier[40](65,45)(69,55)(60,55)
\qbezier[40](105,45)(101,55)(110,55)
\qbezier[40](115,45)(119,55)(110,55)
\qbezier[40](245,45)(241,55)(250,55)
\qbezier[40](255,45)(259,55)(250,55)

% signatures
\put(6,15){$I_0$} \put(56,15){$I_1$} \put(106,15){$I_2$}
\put(246,15){$I_k$}

\multiput(30,21)(50,0){3}{$\scriptstyle{2/3}$}
\multiput(220,21)(50,0){2}{$\scriptstyle{2/3}$}
\multiput(54,59)(50,0){2}{$\scriptstyle{1/3}$}
\put(244,59){$\scriptstyle{1/3}$} \put(8,59){$\scriptstyle{1}$}

\put(174,30){\ldots}  \put(314,30){\ldots}

\end{picture}
\caption{Diagram of spectral weight redistribution under the
action of the composition with the covering
$G_2/G_2^T$}\label{pic_spG}
\end{center}
\end{figure}

Assuming that the initial weight distribution has a form
$q_m(0)=1$, $q_i(0)=0$ for all $i \ne m$, that is, all rectangles
are located in $I_m$, then after $t$ steps of compositions with
$G$-type coverings, we obtain a distribution\footnote{Here
$C_t^{k_1,\ldots,k_l}$ stands for the multinomial coefficient
representing the number of ways to select from $t$ elements $l$
groups, with $k_i$ elements in $i$-th group.}
\begin{equation}\label{distG}
 q_{m-k}(t) = \sum_{\begin{array}{c} \scriptstyle k_1+\ldots+k_l \le t \\ \scriptstyle k_1+2k_2+\ldots+lk_l = k \end{array}}
 C_t^{k_1,\ldots,k_l}\alpha_1^{k_1}\cdot\ldots\cdot\alpha_l^{k_l}\alpha_0^{t-(k_1+\ldots+k_l)}
\end{equation}
for $0 \le k < m$. For a component associated with $I_0$, we use a
trivial estimate $q_0(t) \le 1$. By consideration, if $\alpha_i =
0$, then $k_i=0$, and the corresponding factor $\alpha_i^{k_i}$
in~(\ref{distG}) should be replaced by 1. This remark will be
implied in further calculations. The total weight of the
(considered part of the) covering under construction will increase
$\sigma^t(G)$ times in $t$ steps.

Note that $P_G(1/\sqrt{\tau}) = \pi_G(\tau)$, see~(\ref{pi_G}).
Recall that $\pi_G(x) \to \mu_G$ as ${x \to 1}$. Our final choice
of $\tau$ is such that the inequality~(\ref{cond}) remains valid
after replacing $\lambda_F$ by $\lambda$, and $\mu_G$ by
$\pi_G(\tau)$ ($\tau$ should be small enough).

Observe that our choice implies $\alpha_0 \ne 1$, since otherwise
$\pi_G(\tau) \equiv 1$, and the inequality~(\ref{cond}) cannot be
satisfied. On the other hand, $\mu_G < 1$ due to the compactness
of $G$: there should exist rectangles $R \in G$ with $\rho(R)>1$.

{\bf III.} Now we are ready to state the synthesis algorithm,
namely the rule of combination of the coverings $F$ and $G$.
Choose $\gamma$  satisfying the condition
\begin{equation}\label{gamma}
  -\frac1{\lambda} \log_{\tau} \frac{\sigma(G)}{\sigma(F)} < \gamma \le
    - 2 \log_{\tau}\pi_G(\tau).
\end{equation}
Such $\gamma$ does exist, since the inequality between the left
and the right sides of~(\ref{gamma}) is equivalent
to~(\ref{cond}), where $\lambda_F$ is replaced by $\lambda$, and
$\mu_G$ is replaced by $\pi_G(\tau)$ (just apply a base-$\tau$
logarithm to~(\ref{cond}) and divide by $\lambda$).

To construct the required covering of $A^{\otimes n}$, we assign
two sets of rectangles, $\cal F$ and $\cal G$.
\smallskip

\noindent \fbox{%
\parbox{13.4cm}{ \hspace{3.5mm}
$(i)$ Before the start of the algorithm, the set $\cal G$ is empty, and the set
$\cal F$ contains a $1 \times 1$ rectangle
(a trivial covering of the matrix $A^{\otimes 0}$).

\hspace{3.5mm} $(ii)$ Then, perform $n$ similar steps. A step $t$ does the following:

\hspace{3.5mm} --- with the use of a suitable covering $F$ or $F^T$, transform any
rectangle $R \in \cal F$ into $(F/F^T)\otimes
R$;

\hspace{3.5mm} --- with the use of a suitable covering $G$ or
$G^T$, transform any rectangle $R \in \cal G$ into $(G/G^T)\otimes
R$;

\hspace{3.5mm} --- relocate  from $\cal F$ to $\cal G$ all
rectangles $R \in \cal F$ belonging to the sets $I_m$ with $m \ge
\gamma (n-t)$.

\hspace{3.5mm} $(iii)$ By construction, after any $t$ steps, the
set ${\cal F} \cup {\cal G}$ is a covering of the matrix
$A^{\otimes t}$. In the end of the algorithm, the set $\cal F$ is
empty, and $\cal G$ is a covering of $A^{\otimes n}$. }}
\smallskip

Essentially, this is the algorithm~\cite{agp22e}. The parameter
$\gamma$ controls the switching between the two stages of the
algorithm. Next, we are going to prove that $\cal G$ has the
desired complexity.

For any $m$, we have to relocate rectangles belonging to $I_m \cap
\cal F$ on at most $\frac{d}{\gamma}+2 = O(1)$ (consecutive)
steps, namely while $m-d < \gamma (n-t) \le m$, and one more time,
when $m-d \ge \gamma (n-t)$. On the subsequent steps, rectangles
in $I_m$ don't appear.

Let us turn to complexity bounds. The complexity of a rectangle $R
\in I_m$ in relation to its spectral weight is estimated to be
higher, when $m$ is greater. Recall that $w(R) \asymp
\sqrt{\rho(R)} \sigma(R)$. Therefore, an erroneous assignment of a
rectangle to a set $I_m$ with a higher index $m$ (an error to the
right) leads to overestimation of the complexity.

The complexity of the part of the covering $\cal G$ derived from
rectangles that belonged to $I_m \cap \cal F$ at the moment of
relocation is bounded from above as
\begin{multline}\label{sum_comp}
 L_m  \preceq \sigma^n(F)  \nu^m \left(\frac{\sigma(G)}{\sigma(F)}\right)^{\frac{m}{\gamma}} \cdot  \\
 \cdot \left(  1 + \sum_{k_1 + \ldots + k_l \le \frac{m}{\gamma}}
 C_{\frac{m}{\gamma}}^{k_1,\ldots,k_l}\alpha_1^{k_1}\cdot\ldots\cdot\alpha_l^{k_l}\alpha_0^{\frac{m}{\gamma}-(k_1+\ldots+k_l)}
 \tau^{\frac{m-(k_1+2k_2+\ldots+lk_l)}2} \right). %\sum_{0 \le k \le \frac{m}{u}}
\end{multline}
Here the first factor $\sigma^n(F)$ is the expected weight of the
entire covering under the (optimistic) assumption that we apply
only $F$-type coverings. The second factor $\nu^m$ is the upper
bound for the weight portion of the covering~$\cal F$ associated
with $I_m$ at the moment of relocation. The third factor reflects
the weight increase caused by the application of $G$-type
coverings instead of $F/F^T$ on the last $m/\gamma-O(1)$ steps of
the algorithm. In brackets, an additional factor taking into
account the final distribution of the spectral weight is written.
Namely, the term 1 represents the complexity of rectangles from
$I_0$, and under the sum are written the products of the partial
weight portions of rectangles from $I_{m-k}$ with
$k=k_1+2k_2+\ldots+lk_l$ (provided by~(\ref{distG})), and the
estimate $\tau^{\frac{m-k}2}$ of the ratio between the complexity
and the spectral weight of rectangles from $I_{m-k}$. The summands
with $m \le k$ are excess.

Multinomial coefficients satisfy the standard
inequality\footnote{Easily follows by induction on $l$ from the
well-known relation $C_n^k \le 2^{H(k/n)}$.}
\[ C_n^{k_1,\ldots,k_l} \le 2^{nH(k_1/n,\ldots,k_l/n)}, \qquad
\text{where} \]
\[ H(x_1,\ldots,x_l) = - \sum_{i=0}^l x_i\log_2 x_i, \qquad x_0 = 1-\sum_{i=1}^l x_i,   \]
is the binary entropy function defined on the $l$-dimensional
simplex\footnote{On the boundary, the function is defined by
continuity.} ${\mathbb R}_+^l \cap \{x_1+\ldots+x_l \le 1\}$.

It is easy to check that for any $c_i > 0$,
\begin{equation}\label{H+}
 H(x_1,\ldots,x_l) + \sum_{i=1}^l x_i \log_2 c_i \le \log_2\left(1+\sum_{i=1}^l c_i\right).
\end{equation}
Indeed, it immediately follows from the variant of the H\"older's
inequality
\[ \prod_{i=0}^l a_i^{b_i} \le \sum_{i=0}^l a_i b_i  \]
that holds for $a_i,b_i > 0$, and $\sum b_i = 1$, see
e.g.~\cite[Ch.~V]{mpf93e} (just assign $b_i = x_i$, $a_i =
c_i/x_i$, where $c_0=1$, and take a logarithm; in the case $x_i=0$
for some~$i$, evaluate the limit).

By setting $k_i=\frac{x_i m}{\gamma}$ for $i=1,\ldots,l$, and
applying~(\ref{H+}), we obtain
\begin{multline*}
 \log_2 \left( C_{\frac{m}{\gamma}}^{k_1,\ldots,k_l}\alpha_1^{k_1}\cdot\ldots\cdot\alpha_l^{k_l}\alpha_0^{\frac{m}{\gamma}-(k_1+\ldots+k_l)} \tau^{\frac{m-(k_1+2k_2+\ldots+lk_l)}2} \right) \le \\
 \frac{m}{\gamma} \left( H(x_1,\ldots,x_l) + \sum_{i=1}^l x_i \log_2 \frac{\alpha_i}{\alpha_0\tau^{i/2}} + \log_2 \left(\alpha_0\tau^{\gamma /2}\right)\right) \le \\
 \frac{m}{\gamma} \cdot \left( \log_2\left(\sum_{i=0}^l \frac{\alpha_i}{\tau^{i/2}}\right) + \log_2 \left(\tau^{\gamma/2}\right)
                          \right).
\end{multline*}

Thus, we continue (\ref{sum_comp}) as
\[
 L_m \preceq \sigma^n(F) \left( C_0^{\frac{m}{\gamma}} + m^lC_1^{\frac{m}{\gamma}}
\right), \quad \text{where}
\]
\[
 C_0 = \frac{\sigma(G) \nu^{\gamma}}{\sigma(F)}, \qquad
 C_1 = C_0 \cdot P_G\left(1/\sqrt{\tau}\right) \cdot \tau^{\gamma/2}.    %\left(\sum_{i=0}^l \frac{\alpha_i}{\tau^{iu/2}}\right)
\]
Here the power of $C_0$ corresponds to the contribution of
rectangles from $I_0$, and the power of $C_1$ to the contribution
of the remaining rectangles.

Let us check that $C_1 \le C_0 < 1$. Indeed, in the case $\sigma(G)=\sigma(F)$, the inequality $C_0 = \nu^{\gamma}
<1$ holds trivially. In the other case $\sigma(G)>\sigma(F)$, from the left part of~(\ref{gamma}), it follows that
\[ C_0 = \frac{\sigma(G)}{\sigma(F)}\,\tau^{\lambda \gamma} < 1.  \]
Further, the right part of~(\ref{gamma}) implies $\tau^{\gamma/2} \le 1/ \pi_G(\tau)$,
hence $C_1 \le C_0$.

\begin{pnote}
For our example $A=D_4$, choose $\gamma=1/5$. Then $C_0 < 0.99$,
and $C_1 < 0.95$.
\end{pnote}

Finally, we conclude
\[ \LL_2(A^{\otimes n}) \preceq w({\cal G}) = \sum_{m \ge 0} L_m  \preceq
   \sigma^n(F) \sum_{m \ge 0} \left( C_0^{\frac{m}{\gamma}} + m^lC_1^{\frac{m}{\gamma}} \right) \asymp \sigma^n(F). \]
\vskip-8mm \hfill$\blacksquare$

As a reserve for improving the method, one can suggest a more subtle combination of the two
types of coverings involving reverse relocations of rectangles from $\cal G$ to $\cal F$.

\section{Coverings of the Kneser--Sierpinski matrices}

For a matrix $D_r$, $r=2^t$, we propose a covering $F_t$
consisting solely of rectangles of width 1. It generalizes the
examples of Fig.~\ref{pic_cov} (for $t=2$), and from~\cite{agp22e}
(for $t=3$).

We exploit a simple gradient-fashion approach. First, put into $F_t$
a column labeled by $\emptyset$, then add a row labeled by $\emptyset$
from the remaining part of the matrix. Next, we sequentially extract ones
from all columns, and then from all rows labeled by the size-1 subsets of $[t]$.
Then, we do the same with columns and rows labeled by the size-2 subsets, and go on until we reach $t/2$-size labels.
At this point, all ones in $D_r$ are covered.

Let $s(m,k)$ denote the binomial sum
\[ s(m,k) = C_{m}^{k} + C_{m}^{k+1} + \ldots + C_{m}^{m}.  \]

By construction, any rectangle corresponding to a column labeled by a size-$k$ subset has height $s(t-k,k)$,
and a rectangle corresponding to a row labeled by a size-$k$ subset has length $s(t-k,k+1)$.
Hence,
\[ \sigma(F_t) = \sum_{k=0}^{t/2} C_{t}^{k} \left( \sqrt{s(t-k,k)} + \sqrt{s(t-k,k+1)} \right). \]

Direct calculation shows that the quantity $\log_r\sigma(F_t)$ attains its minimum $1.2502...$ when $t=18$,
and, as easy to verify, it tends to the limit $1.259...$ as $t \to \infty$.

As a correcting covering $G_t$, we take a covering of all columns of $D_r$ by individual rectangles,
as shown on Fig.~\ref{pic_cov}. Easy to see that $\sigma(G_t) =
(\sqrt2+1)^t$, and $\mu_{G_t} = \left(\frac{2}{\sqrt2+1}\right)^t$.

With the use of coverings $F_t, G_t$, it is possible to satisfy the conditions of Theorem~\ref{main}
only for $t \le 15$. It can be directly verified that $\sigma(F_{15}) < 442412$, and $\lambda_{F_{15}} < -0.04$.

\begin{corol}
$\SUM_2(D_N) \preceq N^{\log_{2^{15}} \sigma(F_{15})} \prec N^{1.251}$.
\end{corol}

It's a kind of surprise, that coverings of matrices $D_r$ by
width-1 rectangles appear so efficient for the iterative
procedure. However, they are not optimal in terms of spectral
weight. At least starting from $t=7$, one can construct better
coverings via uniting common parts of columns or rows.

\smallskip

The author thanks Stasys Jukna for helpful comments.

\end{document}

\title{Замечание о сложности покрытий для кронекеровых степеней симметрических матриц}
\date{}
\author{И. С. Сергеев\footnote{e-mail: isserg@gmail.com}}

\maketitle

\begin{abstract}
В настоящей заметке исследуется новый метод построения эффективных
покрытий кронекеровых степеней матриц, предложенный
Дж.~Альманом, Ю.~Гуаном и А.~Падаки~\cite{agp22}. Мы предлагаем альтернативное
доказательство результата о сложности
кронекеровых степеней симметрических матриц в более сильной форме.
Как следствие, верхняя оценка
сложности булевых $N\times N$ матриц Кнезера---Серпинского при
вычислении аддитивными схемами глубины 2 улучшена до $O(N^{1.251})$. Данную
работу можно рассматривать как дополнение к~\cite{js13}.
\end{abstract}

\section{Введение}

Кратко напомним необходимые понятия. Более подробное введение в
проблематику см. в~\cite{js13}.

{\it Прямоугольником} размера $a \times b$ называется сплошь
единичная матрица из $a$ строк и $b$ столбцов. Далее в зависимости
от контекста, под прямоугольником мы также будем иногда понимать
матрицу ранга 1, т.е. состоящую из сплошь единичной подматрицы и
нулей на остальных позициях.

%{\it Вес} матрицы определяется как число ее ненулевых элементов,
%обозначается $|A|$. Пусть  --- прямоугольник размера $a \times b$. Таким образом, $|R| = ab$.
{\it Сложностью}\footnote{В работе~\cite{js13} вместо этого
использовался термин {\it вес}. Здесь мы заменяем его на сложность
во избежание путаницы со спектральным весом прямоугольника.}
прямоугольника $R$ размера $a \times b$ назовем сумму длин его
двух сторон и обозначим $w(R) = a+b$. Введем характеристику
<<узости>> прямоугольника как отношение длин большей и меньшей его
сторон, $\rho(R) = \frac{\max(a,\,b)}{\min(a,\,b)}$. {\it
Спектральный вес} прямоугольника определяется как $\sigma(R) =
\sqrt{ab}$.

Множество прямоугольников $F = \{ R_1, \ldots, R_k \}$ называется
{\it покрытием} булевой матрицы $A$, если
\begin{equation}\label{cov}
 A = R_1 + \ldots + R_k.
\end{equation}
(В этой формуле под $R_i$ понимаются матрицы ранга 1.) Если
операция <<+>> в формуле~(\ref{cov}) является целочисленным
сложением, то покрытие $F$ называем $\SUM$-покрытием; если
дизъюнкцией, то $\OR$-покрытием; если сложением по модулю 2, то
$\XOR$-покрытием.

Сложность покрытия определяется как $w(F) = w(R_1) + \ldots +
w(R_k)$, а спектральный вес --- как $\sigma(F) = \sigma(R_1) +
\ldots + \sigma(R_k)$. {\it $\LL$-cложность} матрицы $A$
определяется при $\LL \in \{\SUM, \OR, \XOR \}$ как минимальная
сложность ее $\LL$-покрытия и обозначается $\LL_2(A)$ (что
означает сложность вычисления матрицы линейными схемами
соответствующего типа с глубиной 2).

\begin{pnote}
Указанные определения расширяются на случай матриц над
произвольным полукольцом $S$. Прямоугольник над $S$ имеет вид
$(c_1, \ldots, c_a)^T \cdot (d_1, \ldots, d_b)$, где $c_i, d_j \in
S \setminus \{0\}$. Покрытие матрицы определяется условием
\[  A = e_1 R_1 + \ldots + e_k R_k, \qquad e_i \in S.  \]
Излагаемые далее результаты распространяются и на вводимую
подобным образом сложность вычисления матриц алгебраическими
линейными схемами глубины 2.
\end{pnote}

Пусть $\sigma(A)$ обозначает минимальный спектральный вес покрытия
матрицы $A$. В силу $w(R) \ge 2\sigma(R)$ для любого
прямоугольника $R$, спектральный вес служит простой нижней оценкой
сложности\footnote{Здесь и далее символы $\asymp$, $\prec$,
$\preceq$ означают соответственно равенство, строгое и нестрогое
неравенства порядков роста.}: $\LL_2(A) \succeq \sigma(A)$.

Удобным свойством спектрального веса является мультипликативность
относительно кронекерова произведения. Напомним, что {\it
кронекеровым произведением} булевых матриц $A$ и $B$ называется
матрица $A \otimes B$, получаемая подстановкой матриц $B$ на
позиции единиц матрицы $A$ и сплошь нулевых матриц такого же
размера~--- на позиции нулей.

Заметим, что если $F$ и $G$ --- покрытия матриц $A$ и $B$, то $F
\otimes G$ --- покрытие\footnote{Кронекерово произведение множеств матриц определяется
 как $F \otimes G = \{ R \otimes R' \mid R \in F, \, R' \in G \}$.} матрицы $A \otimes B$, причем $\sigma(F
\otimes G) = \sigma(F)\sigma(G)$. В частности, если мы строим
покрытие матрицы $A^{\otimes n}$ (кронекеровой степени матрицы
$A$) указанным методом композиции, используя подходящие покрытия
матрицы $A$, то сложность построенного покрытия $H$ будет
удовлетворять оценке $w(H) \succeq \sigma(H) \succeq \sigma^n(A)$.

В работе~\cite{agp22} фактически поставлен вопрос, можем ли мы
получать верхние оценки вида  $\LL_2(A^{\otimes n}) \preceq
\sigma^{n+o(n)}(A)$ или хотя бы $\LL_2(A^{\otimes n}) \preceq
\sigma^{n+o(n)}(F)$ для некоторых подходящих покрытий $F$ матрицы
$A$. То, что это не всегда возможно, показывает пример $A = \begin{bmatrix} 1 & 1 \\ 0 & 0 \end{bmatrix}$.

В общем случае препятствием к получению подобных оценок является
растущая несимметричность прямоугольников из покрытия: сложность
прямоугольника $R$ связана со спектральным весом
соотношением\footnote{Более точно, $w(R) = \left(\sqrt{\rho(R)} +\sqrt{1/\rho(R)}\right)\sigma(R)$.}
$w(R) \asymp \sqrt{\rho(R)}\sigma(R)$. Тем не менее, в нескольких
ситуациях авторам~\cite{agp22} удалось преодолеть указанное
затруднение и получить желаемые оценки: во-первых, в случае, когда
покрытие $F$ удовлетворяет определенным критериям асимметричности,
и во-вторых, когда матрица $A$ симметрическая, а покрытие
одностороннее (все прямоугольники вытянуты в одном направлении).

Пожалуй, наиболее интересным (среди булевых матриц) объектом
приложения теории являются матрицы Кнезера---Серпинского (матрицы
непересечений). Напомним, что булева $N \times N$ {\it матрица
Кнезера---Серпинского} $D_N$ определяется при $N=2^n$ следующим
образом. Нумеруя строки и столбцы $D_N$ всевозможными
подмножествами $u \subset [n]$, полагаем $D[u,v] = 1$ в том и только
том случае, когда $u \cap v = \emptyset$. Кроме того, матрицы
$D_N$ могут быть определены как кронекеровы произведения
\begin{equation}\label{D}
D_N = D_2^{\otimes n} = \underbrace{D_2 \otimes \ldots \otimes
D_2}_n,  \qquad
   D_2 = \begin{bmatrix} 1 & 1 \\ 1 & 0 \end{bmatrix}.
\end{equation}

Вопрос о сложности $\OR$-покрытий матриц $D_N$ практически закрыт
в работе~\cite{cils17}. В результате установлено, что
\[ N^{1.16} \prec \OR_2(D_N) \prec N^{1.17} \]
(нижняя оценка из~\cite{js13}), причем авторы~\cite{cils17}
предъявили покрытие, сложность которого оптимальна с точностью до
множителя $(\log N)^{O(1)}$.

Меньше определенности в вопросе об аддитивной ($\SUM$) сложности матриц. В
работе~\cite{js13} предложен простой способ доказательства верхней
оценки
$$\SUM_2(D_N) \preceq \sigma^n(D_2) = \left(\sqrt2+1\right)^n \prec N^{1.272},$$
который опирается на тривиальное (и оптимальное) покрытие матрицы
$D_2$ прямоугольниками размера $1 \times 1$ и $1\times 2$ (или
$2\times 1$). Этот способ нетривиально обобщен в
работе~\cite{agp22}. В силу присущих анализу метода ограничений
эффективная реализация матриц Кнезера---Серпинского обоснована
только из базовых покрытий матриц $D_4$ и $D_8$. Для последнего
случая в~\cite{agp22} получена оценка
$$\SUM_2(D_N) \preceq \sigma^{n/3+o(n)}(D_8) = \left(\sqrt8+\sqrt7+3\sqrt3+3\right)^{n/3+o(n)} \prec N^{1.258}.$$

В настоящей заметке описан вариант этого метода для (наиболее
интересного) случая симметрических матриц. Границы применимости
(относительно общего метода~\cite{agp22}) расширены и дано более
элементарное доказательство. При этом верхняя оценка сложности матриц
Кнезера---Серпинского уточнена до $\SUM_2(D_N) \prec N^{1.251}$.

По-прежнему открыт вопрос о возможности построения существенно
более эффективных $\XOR$-покрытий для матриц $D_N$.

\section{Метод синтеза}

В этом разделе мы изложим общий метод построения покрытия
симметрической матрицы $A^{\otimes n}$, по существу совпадающий с
методом~\cite{agp22}. По ходу изложение будет иллюстрироваться
простым примером, в котором нетривиальное покрытие матрицы $D_N$
строится при помощи покрытий матрицы $D_4$. Этот пример не требует
полной общности метода.

Пусть в нашем распоряжении есть два покрытия симметрической
матрицы $A$ (точнее, две пары покрытий, если считать вместе с
транспонированными). Первое покрытие $F$ ($F^T$)  эффективно с
точки зрения спектрального веса. Второе покрытие $G$
одностороннее: все его прямоугольники вытянуты в одну сторону.
Покрытия $G/G^T$ будут использоваться для коррекции дисбаланса,
вызванного применением покрытий $F / F^T$.

\begin{pnote}
На рис.~\ref{pic_cov} изображены подходящие покрытия матрицы
$D_4$. Покрытие $F_2$ (слева) имеет оптимальный спектральный вес
$\sigma(F_2) = 4+\sqrt3$. Покрытие $G_2$ (справа) обладает
хорошими корректирующими характеристиками. Его спектральный вес
чуть выше, $\sigma(G_2)=3+2\sqrt2$.
\end{pnote}

\begin{figure}[htb]
\begin{center}
\begin{picture}(280,100)(0,0)

\thicklines

\multiput(0,20)(20,0){2}{\line(0,1){80}}
\multiput(0,20)(0,80){2}{\line(1,0){20}}
\multiput(20,80)(0,20){2}{\line(1,0){60}}
\multiput(40,40)(20,20){3}{\line(0,1){20}}
\put(20,40){\line(1,0){20}} \put(20,60){\line(1,0){40}}
\put(40,60){\line(0,1){20}}

\multiput(7,26)(0,20){4}{$1$} \multiput(27,86)(20,0){3}{$1$}
\multiput(27,46)(20,20){2}{$1$}

\multiput(27,26)(20,0){3}{$0$} \multiput(47,46)(20,0){2}{$0$}
\multiput(27,66)(40,0){2}{$0$}

\put(35,0){$F_2$}

\multiput(200,20)(20,0){2}{\line(0,1){80}}
\multiput(200,20)(20,20){4}{\line(1,0){20}}
\put(200,100){\line(1,0){80}} \put(240,40){\line(0,1){60}}
\put(260,60){\line(0,1){40}} \put(280,80){\line(0,1){20}}

\multiput(207,26)(0,20){4}{$1$} \multiput(227,86)(20,0){3}{$1$}
\multiput(227,46)(20,20){2}{$1$}

\multiput(227,26)(20,0){3}{$0$} \multiput(247,46)(20,0){2}{$0$}
\multiput(227,66)(40,0){2}{$0$}

\put(235,0){$G_2$}

\thinlines

\multiput(223,60)(0,20){2}{\line(1,0){14}}
%\put(220,60){\line(1,1){20}} \put(220,80){\line(1,-1){20}}

\multiput(23,23)(0,7){3}{\circle*{1}}
\multiput(37,23)(0,7){3}{\circle*{1}}
\multiput(30,23)(0,14){2}{\circle*{1}}
\multiput(23,63)(0,7){3}{\circle*{1}}
\multiput(37,63)(0,7){3}{\circle*{1}}
\multiput(30,63)(0,14){2}{\circle*{1}}
\multiput(43,23)(0,7){3}{\circle*{1}}
\multiput(57,23)(0,7){3}{\circle*{1}}
\multiput(50,23)(0,14){2}{\circle*{1}}
\multiput(43,43)(0,7){3}{\circle*{1}}
\multiput(57,43)(0,7){3}{\circle*{1}}
\multiput(50,43)(0,14){2}{\circle*{1}}
\multiput(63,43)(0,7){3}{\circle*{1}}
\multiput(77,43)(0,7){3}{\circle*{1}}
\multiput(70,43)(0,14){2}{\circle*{1}}
\multiput(63,23)(0,7){3}{\circle*{1}}
\multiput(77,23)(0,7){3}{\circle*{1}}
\multiput(70,23)(0,14){2}{\circle*{1}}
\multiput(63,63)(0,7){3}{\circle*{1}}
\multiput(77,63)(0,7){3}{\circle*{1}}
\multiput(70,63)(0,14){2}{\circle*{1}}

\multiput(223,23)(0,7){3}{\circle*{1}}
\multiput(237,23)(0,7){3}{\circle*{1}}
\multiput(230,23)(0,14){2}{\circle*{1}}
\multiput(223,63)(0,7){3}{\circle*{1}}
\multiput(237,63)(0,7){3}{\circle*{1}}
\multiput(230,63)(0,14){2}{\circle*{1}}
\multiput(243,23)(0,7){3}{\circle*{1}}
\multiput(257,23)(0,7){3}{\circle*{1}}
\multiput(250,23)(0,14){2}{\circle*{1}}
\multiput(243,43)(0,7){3}{\circle*{1}}
\multiput(257,43)(0,7){3}{\circle*{1}}
\multiput(250,43)(0,14){2}{\circle*{1}}
\multiput(263,43)(0,7){3}{\circle*{1}}
\multiput(277,43)(0,7){3}{\circle*{1}}
\multiput(270,43)(0,14){2}{\circle*{1}}
\multiput(263,23)(0,7){3}{\circle*{1}}
\multiput(277,23)(0,7){3}{\circle*{1}}
\multiput(270,23)(0,14){2}{\circle*{1}}
\multiput(263,63)(0,7){3}{\circle*{1}}
\multiput(277,63)(0,7){3}{\circle*{1}}
\multiput(270,63)(0,14){2}{\circle*{1}}

\end{picture}
\caption{Оптимальное и компенсирующее покрытия матрицы
$D_4$}\label{pic_cov}
\end{center}
\end{figure}

Пусть $H$ --- покрытие матрицы $B$. Тогда покрытие матрицы
$A \otimes B$ можно получить в виде $\{ F_i \otimes R_i \mid R_i
\in H \}$, где $F_i$ --- некоторые покрытия матрицы $A$. Так мы
последовательно строим покрытия матриц $A, A^{\otimes2},
A^{\otimes3}, \ldots$ В основном процессе мы выбираем $F_i \in \{
F,\, F^T \}$. Более конкретно, к $a \times b$ прямоугольнику $R$
применяется $F$, если $a \le b$, и $F^T$~--- иначе.

Пусть покрытие $F = \{ R_1, \ldots, R_s \}$ %$r \times r$ матрицы $A$
состоит из $a_i \times b_i$ прямоугольников~$R_i$. {\it
Характеристическая функция покрытия} определяется как
\begin{equation}\label{char}
    \chi_F(x) = \sigma(R_1)\left( \frac{a_1}{b_1} \right)^x + \ldots + \sigma(R_s)\left( \frac{a_s}{b_s} \right)^x - \sigma(F).
\end{equation}
В силу определения, $\chi_F(0)=0$. Покрытие $F$ назовем {\it
компактным}, если $\chi_F(x)$ принимает отрицательные значения на
отрицательной полуоси\footnote{Это слабый аналог определения
несбалансированного покрытия в~\cite{agp22}.}. Для компактного
покрытия $F$ обозначим через $\lambda_F$ минимальный вещественный
корень функции $\chi_F$, в правой полуокрестности которого функция
отрицательна\footnote{Из компактности покрытия следует, что $b_i <
a_i$ при некотором $i$. Тогда число $\lambda_F$ определено
корректно ввиду того, что $\chi_F(x) \to +\infty$ при $x \to
-\infty$, а любая экспоненциальная сумма вида~(\ref{char}) имеет
конечное число вещественных нулей, см., например,~\cite{lan31}.}.

Заметим, что условие $\chi'_F(0) = \sum_i \sigma(R_i) \ln(a_i/b_i)
> 0$ является достаточным для компактности покрытия $F$.

Мы требуем, чтобы используемое для вычисления кронекеровой степени
матрицы $A$ покрытие $F$ было компактным.

\begin{pnote}
Покрытие $F_2$ матрицы $D_4$ является компактным. Его
характеристическая функция имеет вид $\chi_{F_2}(x) = 2\cdot4^x +
\sqrt3\cdot3^{-x} - 2- \sqrt{3}$, a минимальный корень равен
$\lambda_{F_2} \approx -0.305$.
\end{pnote}

В роли компенсирующего будет выступать компактное {\it
одностороннее} покрытие $G$, состоящее %  = \{ R'_1, \ldots, R'_l \}
из $a'_i \times b'_i$ прямоугольников $R'_i$, где $a'_i \ge b'_i$.
Качество такого покрытия будет характеризоваться коэффициентом
\[  \mu_G = \frac1{\sigma(G)} \sum_{R \in G} \frac{\sigma(R)}{\sqrt{\rho(R)}} = \frac1{\sigma(G)} \sum_{i} b'_i.  \]

При $x>1$ определим функцию
\begin{equation}\label{pi_G}
  \pi_G(x) = \frac1{\sigma(G)} \sum_{R \in G} \sigma(R)\cdot x^{-\lfloor \log_x \rho(R) \rfloor/2}.
\end{equation}
Из определения вытекает, что $\pi_G(x) \ge \mu_G$ и $\pi_G(x) \to \mu_G$ при $x \to 1$.

\begin{pnote}
Для покрытия $G_2$ справедливо $\mu_{G_2} = \frac4{3+2\sqrt2}$.
%Покрытие $G_2$ является $\left( \frac13;\,4\right)$-разделяющим, тогда $\pi_{G_2}(x) = (x+2)/3$.
\end{pnote}

\begin{theorem}\label{main}
Пусть $F$ --- компактное $\LL$-покрытие, а $G$ ---
компактное одностороннее $\LL$-покрытие симметрической
$r\times r$ матрицы $A$, при этом $\sigma(G) \ge \sigma(F)$. Если
выполнено условие
\begin{equation}\label{cond}
   \frac{\sigma(G)}{\sigma(F)} < \mu_G^{2\lambda_F},
\end{equation}
то для $N=r^n$ справедливо
\[\LL_2(A^{\otimes n}) \preceq N^{\log_r \sigma(F)}.\]
\end{theorem}

\begin{pnote}
Для матрицы $D_4$ и покрытий $F_2, G_2$ условия теоремы выполнены,
т.к. $\lambda_{F_2} < -0.3$ и $\mu_{G_2}= \frac4{3+2\sqrt2}$.
Поэтому $\SUM_2(D_N) \preceq N^{\log_4(4+\sqrt3)} \prec N^{1.26}.$
\end{pnote}

\proof План доказательства следующий. Сначала выполним анализ
эволюции размеров прямоугольников покрытия при многократном
применении покрытий $F/F^T$. Затем --- то же самое для покрытий
$G/G^T$. В заключение опишем подходящую комбинацию двух типов
покрытий.

{\bf I.}  По условию компактности покрытия $F$, для некоторых
(достаточно малых) $\delta, \epsilon >0$ и $\lambda =
\lambda_F+\delta$ выполнено $\chi_F(\lambda) < - \epsilon$ и
одновременно неравенство (\ref{cond}) сохраняет силу при
подстановке $\lambda$ вместо $\lambda_F$. Покажем, что для всех
достаточно малых $\tau>1$ выполняется
\begin{equation}\label{chi_app}
    \sigma(R_1)\tau^{\lambda \lfloor \log_{\tau}(a_1/b_1) \rfloor} + \ldots +
    \sigma(R_s)\tau^{\lambda \lfloor \log_{\tau}(a_s/b_s) \rfloor} \le \sigma(F).
\end{equation}
Действительно, левая часть (\ref{chi_app}) не превосходит
\[ \tau^{-\lambda}(\chi_F(\lambda)+\sigma(F)) < \tau^{-\lambda}(\sigma(F)-\epsilon), \]
поэтому для выполнения (\ref{chi_app}) достаточно потребовать
$\tau^{-\lambda} \le 1 + \frac{\epsilon}{\sigma(F)}$, таким
образом, подходит интервал значений  $1 < \tau \le \left( 1 +
\frac{\epsilon}{\sigma(F)} \right)^{-1/\lambda}$.

Параметру $\tau$ мы припишем смысл шага изменения соотношения
между сторонами при классификации прямоугольников. Окончательный
выбор параметра будет сделан позже.

На множестве прямоугольников $\cal R$ введем классификацию в
зависимости от отношения большей стороны к меньшей: ${\cal R} =
\bigcup_{k \ge 0} I_k$, где $I_0$ содержит прямоугольники $R$, для
которых $\rho(R)\le r$, а при $k \ge 1$ множество~$I_k$ содержит
прямоугольники с отношением сторон $r\cdot\tau^{k-1} < \rho(R)\le
r\cdot\tau^{k}$.  Напомним, что $r$ --- это размер матрицы $A$.

\begin{pnote}
В случае покрытия $F_2$ полагаем $\tau = 4$. Тогда $I_0 = \{ R
\mid \rho(R) \le 4 \}$ и $I_k = \{ R \mid 4^k < \rho(R) \le
4^{k+1} \}$ при $k>0$. Функция $\chi_{F_2}(x)$ принимает
отрицательные значения на всем интервале $(\lambda_{F_2},\, 0)$,
поэтому мы достаточно свободны в выборе $\lambda$. Конкретное
значение будет указано позже.
\end{pnote}

Выполняя композиции с покрытиями матрицы $A$, мы будем отслеживать
перераспределение спектрального веса множества прямоугольников
между множествами $I_k$. При этом мы будем руководствоваться {\it
принципом <<ошибки вправо>>}. Это значит, что: (а) мы допускаем
ошибочное отнесение прямоугольника к множеству $I_k$ с более
высоким индексом $k$, но не наоборот; (б) оценка распределения
веса покрытия $A \otimes R$ выполняется только на основании
принадлежности прямоугольника $R$ тому или иному множеству $I_k$.
Как следствие, распределение спектрального веса покрытия,
полученное в результате серии шагов оценочным образом (с
возможными ошибками вправо), будет отличаться от истинного
распределения только тем, что некоторые прямоугольники будут
ошибочно помещены в множества $I_k$ с б\'ольшими индексами.

Перераспределение спектрального веса множества прямоугольников при
композиции с покрытиями $F/F^T$ описывается (с возможной ошибкой
вправо) коэффициентами многочлена Лорана, получаемого
из~(\ref{chi_app}):
\begin{equation}\label{laur}
   P_F(x) = \sum_{i \in \mathbb Z} \beta_i x^i = \frac{\sigma(R_1)}{\sigma(F)} x^{\lfloor \log_{\tau}(a_1/b_1) \rfloor} + \ldots +
    \frac{\sigma(R_s)}{\sigma(F)} x^{\lfloor \log_{\tau}(a_s/b_s) \rfloor}.
\end{equation}
Это значит, что для $R \in I_m$ (спектральный) вес прямоугольников
из $(F/F^T)\otimes R$ распределяется так: доля $\beta_k$
принадлежит множеству $I_{m-k}$ в случае $m-k>0$, и множеству
$I_0$, иначе.

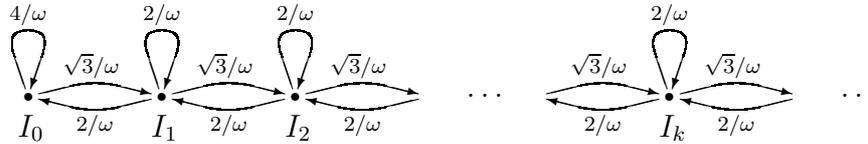
\begin{figure}[htb]
\begin{center}
\begin{picture}(320,60)(0,15)

\multiput(10,30)(50,0){3}{\circle*{3}} \put(250,30){\circle*{3}}
% right arrows
\multiput(44,34)(50,0){3}{\vector(4,-1){12}}
\multiput(14,31)(50,0){3}{\line(4,1){12}}
\multiput(234,34)(50,0){2}{\vector(4,-1){12}}
\multiput(204,31)(50,0){2}{\line(4,1){12}}
\qbezier[40](26,34)(35,36)(44,34)
\qbezier[40](76,34)(85,36)(94,34)
\qbezier[40](126,34)(135,36)(144,34)
\qbezier[40](216,34)(225,36)(234,34)
\qbezier[40](266,34)(275,36)(284,34)

% left arrows
\multiput(26,26)(50,0){3}{\vector(-4,1){12}}
\multiput(56,29)(50,0){3}{\line(-4,-1){12}}
\multiput(216,26)(50,0){2}{\vector(-4,1){12}}
\multiput(246,29)(50,0){2}{\line(-4,-1){12}}
\qbezier[40](26,26)(35,24)(44,26)
\qbezier[40](76,26)(85,24)(94,26)
\qbezier[40](126,26)(135,24)(144,26)
\qbezier[40](216,26)(225,24)(234,26)
\qbezier[40](266,26)(275,24)(284,26)

% cycles
\multiput(15,45)(50,0){3}{\vector(-1,-3){4}}
\put(255,45){\vector(-1,-3){4}}
\multiput(9,33)(50,0){3}{\line(-1,3){4}}
\put(249,33){\line(-1,3){4}}%
\qbezier[40](5,45)(1,55)(10,55)%
\qbezier[40](15,45)(19,55)(10,55)
\qbezier[40](55,45)(51,55)(60,55)
\qbezier[40](65,45)(69,55)(60,55)
\qbezier[40](105,45)(101,55)(110,55)
\qbezier[40](115,45)(119,55)(110,55)
\qbezier[40](245,45)(241,55)(250,55)
\qbezier[40](255,45)(259,55)(250,55)

% signatures
\put(6,15){$I_0$} \put(56,15){$I_1$} \put(106,15){$I_2$}
\put(246,15){$I_k$}

\multiput(28,17)(50,0){3}{$\scriptstyle{2/\omega}$}
\multiput(218,17)(50,0){2}{$\scriptstyle{2/\omega}$}
\multiput(23,39)(50,0){3}{$\scriptstyle{\sqrt3/\omega}$}
\multiput(213,39)(50,0){2}{$\scriptstyle{\sqrt3/\omega}$}
\multiput(53,59)(50,0){2}{$\scriptstyle{2/\omega}$}
\put(243,59){$\scriptstyle{2/\omega}$}
\put(3,59){$\scriptstyle{4/\omega}$}

\put(174,30){\ldots}  \put(314,30){\ldots}

\end{picture}
\caption{Диаграмма перераспределения спектрального веса при
композиции с покрытием $F_2/F_2^T$ (здесь
$\omega=\sigma(F_2)=4+\sqrt3$)}\label{pic_spF}
\end{center}
\end{figure}

\begin{pnote}
Для покрытия $F_2$ при выборе $\tau=4$ получаем $P_{F_2}(x) =
\frac{2x + 2 + \sqrt3x^{-1} }{4+\sqrt3}$. Диаграмма переходов
изображена на рис.~\ref{pic_spF}.
\end{pnote}

Пусть $p_k(t)$ означает долю спектрального веса построенного
покрытия матрицы $A^{\otimes t}$, относящуюся к множеству $I_k$. В
начале имеем $p_0(0)= 1$ и $p_k(0)=0$ при $k>0$. Положим
$\nu=\tau^{\lambda}$. Ввиду~(\ref{chi_app}), $P_F(\nu)\le1$.
Обозначим $d = \deg P_F = \max_{\beta_i >0} |i|$.

Покажем, что распределение $\{ p_k^*(t) \}$, где $p^*_k(t) =
\nu^k$ при $k \le dt$ и $p^*_k(t) = 0$ при $k > dt$, является
мажорантой для $\{ p_k(t) \}$, т.е. величины $p^*_k(t)$ являются
верхними оценками для компонент некоторого распределения
$\{p'_k(t)\}$, получаемого из $\{ p_k(t) \}$ сдвигом части
распределения вправо (из компонент с меньшими индексами в
компоненты с б\'ольшими индексами)\footnote{Хотя распределение $\{
p_k(t) \}$ является вероятностным, от мажоранты $\{ p_k^*(t) \}$
мы этого не требуем, допуская чтобы сумма компонент распределения
была больше 1. Нашей задачей является лишь получение верхних
оценок на $p_k(t)$.}.

Очевидно, в момент $t=0$ условие мажорирования выполнено. Проверим
индуктивный переход: применим композицию с $F/F^T$ к множеству
прямоугольников с распределением $\{ p_k^*(t)\}$. Прямоугольников
из $I_k$, где $k > d(t+1)$, при этом не возникает. В случае $0<k
\le d(t+1)$ вес группы прямоугольников из $I_k$ после
преобразования оценивается сверху как
\begin{equation}\label{w_bal}
     \sum_{i \in \mathbb Z} \beta_i \nu^{k+i} = \nu^k P_F(\nu) \le \nu^k.
\end{equation}
В случае $k=0$ такая оценка, вообще говоря, неверна, т.к. в $I_0$
дополнительно остается значительная часть веса.

Однако, если $I_0$ получает избыток веса, то множества $I_1, I_2,
\ldots$ (в общем случае) имеют его дефицит, как следует
из~(\ref{w_bal}). Таким образом, мы можем перераспределить
избыточный вес из $I_0$ в другие множества, допуская ошибку
вправо. Это возможно, поскольку относительный суммарный вес (сумма
компонент распределения) сохраняется при композиции.

Как следствие, после $t$ шагов композиции с покрытиями $F/F^T$, в
множестве $I_k$ сосредоточено не более $\nu^k$ от общего веса
$\sigma^t(F)$ построенного покрытия матрицы $A^{\otimes t}$ (с
точностью до ошибок вправо).

\begin{pnote}
В примере с матрицей $D_4$ мы можем выбрать $\nu = \sqrt3/2$ (это
корень многочлена $P_{F_2}(x)$). Такой выбор позволяет не
прибегать к перераспределению веса из~$I_0$. Распределение $\{1,
\nu, \nu^2, \ldots, \nu^k, \ldots\}$ является стационарным для
диаграммы рис.~\ref{pic_spF}. Косвенным образом мы выбрали
$\lambda = \log_4\nu > \lambda_{F_2}$.
\end{pnote}

{\bf II.} Перераспределение веса при композиции с покрытиями
$G/G^T$ описывается (с допустимой ошибкой вправо) многочленом
\[ P_G(x) = \alpha_l x^{l} + \ldots + \alpha_1 x + \alpha_0 = \frac1{\sigma(G)} \sum_{R \in G} \sigma(R)x^{\lfloor \log_{\tau} \rho(R) \rfloor}. \]
Для $R \in I_m$ спектральный вес прямоугольников из $(G/G^T)\otimes R$
распределяется так: доля $\alpha_k$ принадлежит множеству $I_{m-k}$
в случае $k<m$, и множеству $I_0$, иначе.

\begin{figure}[htb]
\begin{center}
\begin{picture}(320,60)(0,15)

\multiput(10,30)(50,0){3}{\circle*{3}} \put(250,30){\circle*{3}}

% left arrows
\multiput(57,30)(50,0){3}{\vector(-1,0){44}}
\multiput(247,30)(50,0){2}{\vector(-1,0){44}}

% cycles
\multiput(15,45)(50,0){3}{\vector(-1,-3){4}}
\put(255,45){\vector(-1,-3){4}}
\multiput(9,33)(50,0){3}{\line(-1,3){4}}
\put(249,33){\line(-1,3){4}}%
\qbezier[40](5,45)(1,55)(10,55)%
\qbezier[40](15,45)(19,55)(10,55)
\qbezier[40](55,45)(51,55)(60,55)
\qbezier[40](65,45)(69,55)(60,55)
\qbezier[40](105,45)(101,55)(110,55)
\qbezier[40](115,45)(119,55)(110,55)
\qbezier[40](245,45)(241,55)(250,55)
\qbezier[40](255,45)(259,55)(250,55)

% signatures
\put(6,15){$I_0$} \put(56,15){$I_1$} \put(106,15){$I_2$}
\put(246,15){$I_k$}

\multiput(30,21)(50,0){3}{$\scriptstyle{2/3}$}
\multiput(220,21)(50,0){2}{$\scriptstyle{2/3}$}
\multiput(54,59)(50,0){2}{$\scriptstyle{1/3}$}
\put(244,59){$\scriptstyle{1/3}$} \put(8,59){$\scriptstyle{1}$}

\put(174,30){\ldots}  \put(314,30){\ldots}

\end{picture}
\caption{Диаграмма перераспределения спектрального веса при
композиции с покрытием $G_2/G_2^T$}\label{pic_spG}
\end{center}
\end{figure}
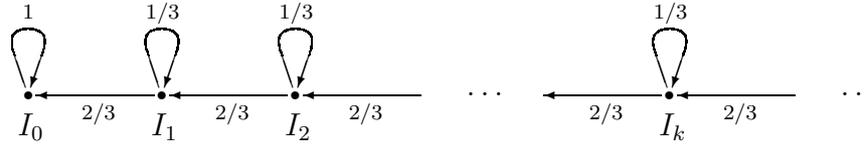

\begin{pnote}
Для покрытия $G_2$ при $\tau=4$ можно выбрать (с ошибкой вправо) $P_{G_2}(x) = (x+2)/3$.
Перераспределение веса под
действием покрытий $G_2/G_2^T$ описывается диаграммой
рис.~\ref{pic_spG}.
\end{pnote}

Таким образом, если в начальный момент времени распределение
спектрального веса записывалось как $q_m(0)=1$, $q_i(0)=0$ при
всех $i \ne m$ (т.е. все прямоугольники находятся в $I_m$), то
после $t$ итераций композиции с покрытиями $G/G^T$ распределение
примет вид\footnote{Обозначение $C_t^{k_1,\ldots,k_l}$ используется для
мультиномиального коэффициента --- это число способов выбрать из $t$ элементов $l$ групп,
в $i$-й группе~--- $k_i$ элементов.}
\begin{equation}\label{distG}
 q_{m-k}(t) = \sum_{\begin{array}{c} \scriptstyle k_1+\ldots+k_l \le t \\ \scriptstyle k_1+2k_2+\ldots+lk_l = k \end{array}}
 C_t^{k_1,\ldots,k_l}\alpha_1^{k_1}\cdot\ldots\cdot\alpha_l^{k_l}\alpha_0^{t-(k_1+\ldots+k_l)},
\end{equation}
при $0 \le k < m$. Долю, приходящуюся на $I_0$, оценим просто как
$q_0(t) \le 1$. Примем соглашение, что если $\alpha_i = 0$, то
$k_i=0$, и соответствующий множитель $\alpha_i^{k_i}$ в
(\ref{distG}) равен 1. Это замечание будет подразумеваться в
последующих выкладках. Общий вес (рассматриваемой части) покрытия
за $t$ шагов увеличится в $\sigma^t(G)$ раз.

Заметим, что $P_G(1/\sqrt{\tau}) = \pi_G(\tau)$, см.~(\ref{pi_G}). Напомним, что $\pi_G(x) \to \mu_G$ при $x \to 1$.
Окончательно выбираем параметр $\tau$ (достаточно малым) так,
чтобы неравенство~(\ref{cond}) сохранялось при подстановке
$\lambda$ вместо $\lambda_F$ и $\pi_G(\tau)$ вместо $\mu_G$.

Отметим, что при таком выборе $\alpha_0 \ne 1$, иначе $\pi_G(\tau) \equiv 1$, поэтому
неравенство~(\ref{cond}) не может быть удовлетворено. С другой стороны, $\mu_G < 1$
ввиду компактности покрытия $G$ (найдется прямоугольник $R \in G$, для которого $\rho(R)>1$).

{\bf III.} Теперь опишем сам алгоритм синтеза, а именно, правило
комбинации покрытий $F$ и $G$. Выберем $\gamma$ из условия
\begin{equation}\label{gamma}
  -\frac1{\lambda} \log_{\tau} \frac{\sigma(G)}{\sigma(F)} < \gamma \le
    - 2 \log_{\tau}\pi_G(\tau).
\end{equation}
Такое $\gamma$ существует, т.к. неравенство между левой и правой
частями~(\ref{gamma}) эквивалентно неравенству~(\ref{cond}) с
подстановкой $\lambda$ вместо $\lambda_F$ и $\pi_G(\tau)$ вместо $\mu_G$
(следует взять логарифм по основанию $\tau$ от (\ref{cond}) и разделить на $\lambda$).

Для построения покрытия матрицы $A^{\otimes n}$ отведем два
множества прямоугольников, $\cal F$ и $\cal G$. \smallskip

\noindent \fbox{%
\parbox{13.4cm}{ \hspace{3.5mm}
$(i)$ Перед началом работы алгоритма множество $\cal G$ пусто,
множество $\cal F$ содержит прямоугольник размера $1 \times 1$
(тривиальное покрытие матрицы $A^{\otimes 0}$).

\hspace{3.5mm} $(ii)$ Далее выполняем $n$ однотипных шагов. Шаг
$t$ заключается в следующем:

\hspace{3.5mm} --- используя подходящее покрытие $F$ или $F^T$,
каждый прямоугольник $R \in \cal F$ преобразуем в $(F/F^T)\otimes
R$;

\hspace{3.5mm} --- используя подходящее покрытие $G$ или $G^T$,
каждый прямоугольник $R \in \cal G$ преобразуем в $(G/G^T)\otimes
R$;

\hspace{3.5mm} --- все прямоугольники $R \in \cal F$,
принадлежащие множествам $I_m$ с номерами $m \ge \gamma (n-t)$,
перемещаем из $\cal F$ в $\cal G$.

\hspace{3.5mm} $(iii)$ По построению, после $t$ шагов множество
${\cal F} \cup {\cal G}$ является покрытием матрицы $A^{\otimes
t}$. В конце работы алгоритма множество $\cal F$ пусто, а
множество $\cal G$ является покрытием $A^{\otimes n}$. }}
\smallskip

%Последовательно на протяжении $n$ шагов выполняем композицию всех
%прямоугольников с покрытиями $F/F^T$, но после каждого шага $t$
%прямоугольники, принадлежащие множествам $I_m$ с номерами $m \ge
%\gamma (n-t)$, изымаются из общего покрытия и преобразуются далее
%только при помощи композиций с $G/G^T$ (на оставшихся $n-t$ шагах).

По существу, это и есть алгоритм~\cite{agp22}. Параметр $\gamma$
управляет переключением между двумя стадиями алгоритма. Проверим,
что покрытие~$\cal G$ имеет требуемую сложность.

%Таким образом, прямоугольники из множеств $I_m$ c номерами $m >
%\gamma n+d$ вовсе не появятся по ходу алгоритма.
При любом $m$ прямоугольники из $I_m$  %при каждом $m$ % \le \gamma n+d$
отправляются из $\cal F$ в $\cal G$ не более чем на
$\frac{d}{\gamma}+2 = O(1)$ (последовательных) итерациях, а
именно, пока $m-d < \gamma (n-t) \le m$ и еще один раз, когда $m-d
\ge \gamma (n-t)$. На последующих шагах прямоугольников в $I_m$ не
возникает.

Перейдем к оценкам сложности. Сложность прямоугольника $R \in I_m$
оценивается тем выше относительно спектрального веса, чем больше
$m$: напомним, что $w(R) \asymp \sqrt{\rho(R)} \sigma(R)$. Поэтому
ошибочное отнесение прямоугольника к множеству с более высоким
номером (ошибка вправо) приводит к завышению оценки сложности.

Сложность части покрытия, построенного из прямоугольников,
относившихся к $I_m$ при перемещении в $\cal G$, теперь
оценивается сверху как
\begin{multline}\label{sum_comp}
 L_m  \preceq \sigma^n(F)  \nu^m \left(\frac{\sigma(G)}{\sigma(F)}\right)^{\frac{m}{\gamma}} \cdot  \\
 \cdot \left(  1 + \sum_{k_1 + \ldots + k_l \le \frac{m}{\gamma}}
 C_{\frac{m}{\gamma}}^{k_1,\ldots,k_l}\alpha_1^{k_1}\cdot\ldots\cdot\alpha_l^{k_l}\alpha_0^{\frac{m}{\gamma}-(k_1+\ldots+k_l)}
 \tau^{\frac{m-(k_1+2k_2+\ldots+lk_l)}2} \right). %\sum_{0 \le k \le \frac{m}{u}}
\end{multline}
Здесь первый множитель $\sigma^n(F)$ --- это ожидаемый вес общего
покрытия в предположении выполнения композиции с покрытиями
$F/F^T$. Второй множитель $\nu^m$~--- это оценка доли веса
покрытия, приходящейся на множество $I_m$ на первой стадии
алгоритма (до перемещения в $\cal G$). Третий множитель отвечает
росту веса рассматриваемой части покрытия за счет применения
покрытий типа $G$ вместо $F$ на заключительных $m/\gamma-O(1)$
шагах алгоритма. Допонительный множитель в скобках учитывает
финальное распределение спектрального веса. А именно, слагаемое 1
отвечает сложности прямоугольников из $I_0$, а под знаком суммы
записаны произведения частичных долей веса покрытия после
$m/\gamma$ композиций с $G/G^T$, приходящихся на множество
$I_{m-k}$, где $k=k_1+2k_2+\ldots+lk_l$ (см.~(\ref{distG})), и
оценки $\tau^{\frac{m-k}2}$ отношения сложности прямоугольников из
$I_{m-k}$ к спектральному весу (слагаемые, для которых $m \le k$,
в сумме лишние).

Мультиномиальные коэффициенты удовлетворяют стандартному
неравенству\footnote{Легко проверяется индукцией по $l$, исходя из
широко известного соотношения $C_n^k \le 2^{H(k/n)}$.}
\[ C_n^{k_1,\ldots,k_l} \le 2^{nH(k_1/n,\ldots,k_l/n)}, \qquad \text{где} \]
\[ H(x_1,\ldots,x_l) = - \sum_{i=0}^l x_i\log_2 x_i, \qquad x_0 = 1-\sum_{i=1}^l x_i, \quad \text{---}  \]
определенная на $l$-мерном симплексе ${\mathbb R}_+^l \cap \{x_1+\ldots+x_l \le 1\}$ функция двоичной энтропии\footnote{На границе функция доопределяется по непрерывности.}.

Несложно проверить, что при любых $c_i > 0$ выполнено
\begin{equation}\label{H+}
 H(x_1,\ldots,x_l) + \sum_{i=1}^l x_i \log_2 c_i \le \log_2\left(1+\sum_{i=1}^l c_i\right).
\end{equation}
Это простое следствие варианта неравенства Г\"ельдера
\[ \prod_{i=0}^l a_i^{b_i} \le \sum_{i=0}^l a_i b_i,  \]
справедливого при $a_i,b_i > 0$ и $\sum b_i = 1$,
см.~\cite[Ch.~V]{mpf93} (достаточно подставить $b_i = x_i$, $a_i =
c_i/x_i$, где $c_0=1$, и взять логарифм; в случае $x_i=0$~---
использовать предельный переход).

Полагая $k_i=\frac{x_i m}{\gamma}$ для $i=1,\ldots,l$ и применяя % результат утверждения~\ref{claim}
(\ref{H+}), получаем оценку
\begin{multline*}
 \log_2 \left( C_{\frac{m}{\gamma}}^{k_1,\ldots,k_l}\alpha_1^{k_1}\cdot\ldots\cdot\alpha_l^{k_l}\alpha_0^{\frac{m}{\gamma}-(k_1+\ldots+k_l)} \tau^{\frac{m-(k_1+2k_2+\ldots+lk_l)}2} \right) \le \\
 \frac{m}{\gamma} \left( H(x_1,\ldots,x_l) + \sum_{i=1}^l x_i \log_2 \frac{\alpha_i}{\alpha_0\tau^{i/2}} + \log_2 \left(\alpha_0\tau^{\gamma /2}\right)\right) \le \\
 \frac{m}{\gamma} \cdot \left( \log_2\left(\sum_{i=0}^l \frac{\alpha_i}{\tau^{i/2}}\right) + \log_2 \left(\tau^{\gamma/2}\right)
                          \right).
\end{multline*}

Таким образом,
\[
 L_m \preceq \sigma^n(F) \left( C_0^{\frac{m}{\gamma}} + m^lC_1^{\frac{m}{\gamma}}
\right), \quad \text{где}
\]
\[
 C_0 = \frac{\sigma(G) \nu^{\gamma}}{\sigma(F)}, \qquad
 C_1 = C_0 \cdot P_G\left(1/\sqrt{\tau}\right) \cdot \tau^{\gamma/2}.    %\left(\sum_{i=0}^l \frac{\alpha_i}{\tau^{iu/2}}\right)
\]
Здесь степени $C_0$ отвечают вкладу в сложность прямоугольников из $I_0$, а степени $C_1$ --- вкладу
остальных прямоугольников построенного покрытия.

Проверим, что $C_1 \le C_0 < 1$. Действительно, в случае $\sigma(G)=\sigma(F)$ неравенство $C_0 = \nu^{\gamma}
<1$ выполняется тривиально. В случае $\sigma(G)>\sigma(F)$ из левой части~(\ref{gamma}) вытекает
\[ C_0 = \frac{\sigma(G)}{\sigma(F)}\,\tau^{\lambda \gamma} < 1.  \]
Далее, из правой части~(\ref{gamma}) следует $ \tau^{\gamma/2} \le 1/ \pi_G(\tau)$,
откуда получаем $C_1 \le C_0$.

\begin{pnote}
В примере $A=D_4$ можно выбрать $\gamma=1/5$. Тогда $C_0 < 0.99$
и $C_1 < 0.95$.
\end{pnote}

Окончательно имеем
\[ \LL_2(A^{\otimes n}) \preceq w({\cal G}) =  \sum_{m \ge 0} L_m  \preceq
   \sigma^n(F) \sum_{m \ge 0} \left( C_0^{\frac{m}{\gamma}} + m^lC_1^{\frac{m}{\gamma}} \right) \asymp \sigma^n(F). \]
\vskip-8mm \hfill$\blacksquare$

Резерв для совершенствования метода --- более тонкое комбинирование двух типов покрытий,
с возможностью возврата прямоугольников из $\cal G$ в $\cal F$.

\section{Покрытия матриц Кнезера---Серпинского}

Для произвольной матрицы $D_r$, $r=2^t$, построим компактное
покрытие $F_t$, состоящее исключительно из прямоугольников ширины
1, обобщая примеры рис.~\ref{pic_cov} (для $t=2$) и~\cite{agp22}
(для $t=3$).

Воспользуемся простым вариантом градиентного метода. Сначала
поместим в $F_t$ столбец матрицы $D_r$ с индексом $\emptyset$,
затем --- строку с индексом $\emptyset$ оставшейся части матрицы.
Далее мы последовательно извлекаем единицы из всех столбцов, а
затем из всех строк, помеченных множествами мощности 1. Такую же
процедуру проделаем со столбцами и, затем, со строками,
помеченными множествами мощности 2, и так до тех пор, пока не
дойдем до множеств мощности $t/2$. После этого все единицы матрицы
$D_r$ будут накрыты.

Обозначим через $s(m,k)$ биномиальную сумму
\[ s(m,k) = C_{m}^{k} + C_{m}^{k+1} + \ldots + C_{m}^{m}.  \]

По построению, прямоугольник, соответствующий столбцу с
ин\-дек\-сом-мно\-же\-ством мощности~$k$, имеет высоту $s(t-k,k)$,
а длина прямоугольника, соответствующего строке с
индексом-множеством мощности~$k$, равна $s(t-k,k+1)$. Поэтому
\[ \sigma(F_t) = \sum_{k=0}^{t/2} C_{t}^{k} \left( \sqrt{s(t-k,k)} + \sqrt{s(t-k,k+1)} \right). \]

Непосредственный расчет показывает, что величина $\log_r
\sigma(F_t)$ достигает минимального значения $1.2502...$ при
$t=18$, а при $t \to \infty$, как нетрудно проверить, стремится к пределу $1.259...$

В качестве корректирующего покрытия $G_t$ возьмем покрытие
столбцов матрицы $D_r$ индивидуальными прямоугольниками, как на
рис.~\ref{pic_cov}. Легко видеть, что $\sigma(G_t) =
(\sqrt2+1)^t$ и $\mu_{G_t} = \left(\frac{2}{\sqrt2+1}\right)^t$.

Используя пару покрытий $F_t, G_t$, удовлетворить условиям
теоремы~\ref{main} оказывается возможным только при $t \le 15$.
Непосредственно проверяется, что $\sigma(F_{15}) < 442412$ и $\lambda_{F_{15}} < -0.04$.

\begin{corol}
$\SUM_2(D_N) \preceq N^{\log_{2^{15}} \sigma(F_{15})} \prec N^{1.251}$.
\end{corol}

Несколько неожиданно то, что покрытия матриц $D_r$
прямоугольниками ширины~1 оказываются эффективными для
итерационного процесса. Конечно, они не являются оптимальными по
спектральному весу: по крайней мере, при $t\ge7$ можно строить
лучшие покрытия, объединяя общие части групп строк или столбцов.

\smallskip

Автор признателен Стасису Юкне за полезные комментарии.